\documentstyle[12pt, a4]{article}
\begin{document}
\author{B. G\"{o}n\"{u}l and M. Ko\c{c}ak
\and Department of Engineering Physics, Faculty of Engineering,
\and University of Gaziantep, 27310 Gaziantep -T\"{u}rkiye}
\title{Systematic Search of Exactly Solvable Non-Central Potentials}
\date{}
\maketitle
\begin{abstract}
Recently developed supersymmetric perturbation theory has been
successfully employed to make a complete mathematical analysis of
the reason behind exact solvability of some non-central
potentials. This investigation clarifies once more the
effectiveness of the present formalism.
\end{abstract}



\section{Introduction}

Using the basic ingredient of supersymmetry, a simple alternative
approach has been recently put forward with its application to
perturbed Coulomb potentials \cite{ozer25812003} in arbitrary
dimensions, and subsequently a general procedure within this novel
framework has been given for the exact treatment of quantum states
having nonzero angular momenta \cite{gonul1}, together with other
work \cite{gonul2} in which it has been clearly shown that
approaches based on logarithmic and supersymmetric perturbation
theories are involved within the more general framework of the
formalism.

Gaining confidence from these works we aim with this Letter to
illustrate the idea of this basic but powerful technique can also
be readily used to search exact solvability of non-central
potentials, which clarifies the systematic behind such algebraic
treatments.

Analytically solvable potentials are important for a number of
reasons such as providing model problems to analyze, to start
perturbation theory expansions from, or to provide complete sets
of basis functions for solving real problems. In this respect,
using the ideas of supersymmetry and shape invariance
\cite{cooper2671995}, many authors
\cite{khare10081994}-\cite{kocak21272002} obtained the solutions
of a wide class of non-central potentials in a closed form.
Additionally, in a recent work \cite{hounkonnou18592004} similar
techniques have been used to determine the spectrum of a
vibrational molecular system and using some well-known shape
invariant potentials the authors have obtained energy levels of
triatomic molecules for 12 classes of non-central but separable
potentials. Nevertheless, to the best of our knowledge, the answer
of the natural question that why some of the non-central
potentials can be solved exactly has not been discussed in the
literature, which is the task of the present work.
\section{The Model}
We first start with a brief introduction of the present formalism.
Throughout the paper the unit system $\hbar=2m=1$ is chosen. The
goal in the supersymmetric quantum theory is to solve the Riccati
equation,

\begin{equation}
W^{2}(r)-W'(r)=V(r)-E_{0},
\end{equation}
where $V(r)$ is the potential of interest and $E_{0}$ is the
corresponding ground state energy. If we find $W(r)$, the so
called superpotential, we have of course found the ground state
wave function via,
\begin{equation}
\Psi_{0}(r)=N exp\left[-\int W(z)dz\right],
\end{equation}
 where $N$ is the normalization constant. If $V(r)$ is a shape invariant
 potential,we can in fact obtain the entire spectrum of bound
 state energies and wave functions via the ladder operators \cite{cooper2671995}.
 Now, suppose that we are interested in a potential for which we
 do not know $W(r)$ exactly. More specifically, we assume that
 $V(r)$ differs by a small amount from a potential $V_{0}(r)$ plus
 angular momentum barrier if any, for which one solves the Riccati
 equation explicitly. For the consideration of spherically
 symmetric potentials, the corresponding Schr\"{o}dinger equation for
 the radial wave function has the form
\begin{equation}
\frac{\Psi''_{n}(r)}{\Psi_{n}(r)}=[V(r)-E_{n}],
~~~V(r)=\left[V_{0}(r)+\frac{\ell(\ell+1)}{r^{2}}\right]+\Delta{V(r)}
\end{equation}
 where $\Delta{V}$ is a perturbing potential. Let us write the wave
 function $\Psi_{n}$ as
 \begin{equation}
\Psi_{n}(r)=\chi_{n}(r)\phi_{n}(r),
\end{equation}
 in which $\chi_{n}$ is the known normalized eigenfunction of the
 unperturbed Schr\"{o}dinger equation whereas $\phi_{n}$ is a moderating
 function corresponding to the perturbing potential. Substituting
 (4)into (3) yields
 \begin{equation}
\left(\frac{\chi''_{n}}{\chi_{n}}+\frac{\phi''_{n}}{\phi_{n}}+2\frac{\chi'_{n}}{\chi_{n}}\frac{\phi'_{n}}{\phi_{n}}\right)=V-E_{n}.
\end{equation}
Instead of setting the functions $\chi_{n}$ and $\phi_{n}$, we
will set their logarithmic derivatives using the spirit of Eqs.(1)
and (2);
\begin{equation}
W_{n}=-\frac{\chi'_{n}}{\chi_{n}},
~~~\Delta{W_{n}}=-\frac{\phi'_{n}}{\phi_{n}}
\end{equation}
 which leads to
\begin{equation}
\frac{\chi''_{n}}{\chi_{n}}=W_{n}^{2}-W'_{n}=\left[V_{0}(r)+\frac{\ell(\ell+1)}{r^{2}}\right]-\varepsilon_{n},
\end{equation}
where $\varepsilon_{n}$ is the eigenvalue of the unperturbed and
exactly solvable potential, and
\begin{equation}
\left(\frac{\phi''_{n}}{\phi_{n}}+2\frac{\chi'_{n}}{\chi_{n}}\frac{\phi'_{n}}{\phi_{n}}\right)=
\Delta {W^{2}_{n}}-\Delta {W'_{n}}+2W_{n}\Delta {W_{n}}=\Delta
{V(r)}-\Delta\varepsilon_{n}
\end{equation}
in which $\Delta\varepsilon_{n}$ is the eigenvalue for the
perturbed potential, and
$E_{n}=\varepsilon_{n}+\Delta\varepsilon_{n}$. Then Eq.(5), and
subsequently Eq.(3), reduces to
\begin{equation}
(W_{n}+\Delta {W_{n}})^{2}-(W_{n}+\Delta {W_{n}})'=V-E_{n},
\end{equation}
 which is similar to Eq. (1). In principle as one knows explicitly the
solution of Eq. (7), namely the whole spectrum and corresponding
eigenfunctions of the unperturbed interaction potential, the goal
here is to solve only Eq. (8), which is the backbone of this
formalism, leading to the solution of Eqs. (3) and (9).

Although this is not the case in this paper, if the whole
potential has no analytical solution, which means $\Delta{W}$ and
subsequently Eq. (8) cannot be exactly solvable, then one can
expand the functions  in terms of the perturbation parameter
$\lambda$,
\begin{eqnarray}
\Delta{V(r;\lambda)}=\sum^{\infty}_{k=1}\lambda^{k}\Delta{V_{k}(r)},
~~~\Delta{W_{n}(r;\lambda)}=\sum^{\infty}_{k=1}\lambda^{k}\Delta{W_{nk}(r)},\nonumber
\end{eqnarray}
\begin{eqnarray}
\Delta\varepsilon_{n}(\lambda)=\sum^{\infty}_{k=1}\lambda^{k}\varepsilon_{nk}
\end{eqnarray}
where $\lambda$ will eventually be set equal to one. Substitution
of the above expansion into Eq. (8) by equating terms with the
same power of $\lambda$ on both sides yields up to
$O(\lambda^{3})$
\begin{equation}
2W_{n}\Delta {W_{n1}}-\Delta {W'_{n1}}=\Delta
{V_{1}}-\Delta\varepsilon_{n1},
\end{equation}
\begin{equation}
\Delta{W^{2}_{n1}}+2W_{n}\Delta{W_{n2}}-\Delta{W'_{n2}}=\Delta{V_{2}}-\Delta\varepsilon_{n2}
\end{equation}
\begin{equation}
2(W_{n}\Delta{W_{n3}+\Delta{W_{n1}}\Delta{W_{n2}}})-\Delta{W'_{n3}}=\Delta{V_{3}}-\Delta\varepsilon_{n3},
\end{equation}
Eq. (8) and its expansion, Eqs. (11-13), give a flexibility for
the easy calculations of the perturbative corrections to energy
and wave functions for the   state of interest through an
appropriately chosen perturbed superpotential, unlike the other
perturbation theories.  It has been shown  \cite{gonul2} that this
feature of the present model leads to a simple framework in
obtaining the corrections to all states without using complicated
and tedious mathematical procedures.
\section{Application}
For the consideration of exactly solvable non-central potentials,
\begin{equation}
U(r,\theta,\varphi)=U_{1}(r)+\frac{U_{2}(\theta)}{r^{2}}+\frac{U_{3}(\varphi)}{r^{2}\sin^{2}(\theta)},
\end{equation}
the time-independent Schr\"{o}dinger equation reads
\begin{equation}
\frac{d^{2}R}{dr^{2}}+\frac{2}{r}\frac{dR}{dr}+\left[E-U_{1}-\frac{\ell(\ell+1)}{r^{2}}\right]R=0,
\end{equation}
\begin{equation}
\frac{d^{2}P}{d\theta^{2}}+\cot\theta\frac{dP}{d\theta}+\left[\ell(\ell+1)-\frac{m^{2}}{\sin^{2}(\theta)}-U_{2}\right]P=0
\end{equation}
\begin{equation}
\frac{d^{2}\phi}{d\varphi^{2}}+(m^{2}-U_{3})\phi=0
\end{equation}
where the total wavefunction is
$\Psi(r,\theta,\varphi)=R(r)P(\theta)\phi(\varphi)$  and
$\ell=0,1,2,...,n-1 $ together with $ m=0,\pm1,...,\pm\ell $ are
respectively the orbital and azimuthal quantum numbers. Here, the
crucial point is that each part of the physically total
interaction potential, namely $U_{1}, U_{2}$ and $U_{3}$, should
be analytically solvable. As Eqs. (15) and (17), having an exactly
solvable potential, were well discussed in the literature based in
particular on the supersymmetric quantum theory
\cite{cooper2671995}, we apply the technique presented in the
previous section to only Eq. (16)  to discuss the systematic
behind such equations. However, one should bear in mind that the
same procedure also can be employed easily in Eqs. (15) and (17),
if necessary.

To proceed we use a mapping function $\theta=f(z)$ which
transforms Eq. (16) into
\begin{equation}
\frac{d^{2}P}{dz^{2}}+\left(-\frac{f''}{f'}+f'\cot{f}\right)\frac{dP}{dz}+f'^{2}\left[\ell(\ell+1)-\frac{m^{2}}{\sin^{2}f}-U_{2}(f)\right]P=0.
\end{equation}
The aim here is to have a Schr\"{o}dinger-like equation, therefore
the second term above is removed with the choice of $\theta\equiv
f=2\tan^{-1}(e^{z})$ which yields $\sin\theta=\sec{h}{z},
~\cos\theta=-\tanh{z}$, leading to
\begin{equation}
-\frac{d^{2}P}{dz^{2}}+[U_{2}(z)-\ell(\ell+1)\sec{h}^{2}{z}]P=-m^{2}P
\end{equation}
Now, the question is which forms of $U_{2}$  reproduce analytical
solutions. To answer this question one needs to use the discussion
given by Eqs. (3) through (9). As the whole interaction potential
is,
\begin{equation}
V(z)=V_{0}(z)+\Delta{V(z)}=-\ell(\ell+1)\sec{h}^{2}{z}+U_{2}(z),
\end{equation}
in case the angular part of the potential $U_{2}=0$ in (19), the
remain piece leads to the well-known shape invariant exactly
solvable potential \cite{cooper2671995}. It can be readily solved
by the supersymmetric quantum theory,
\begin{equation}
W_{n=0}(z)=\ell\tanh{z} ,~~~\varepsilon_{n}=-(\ell-n)^{2},
~~~n=0,1,2,...
\end{equation}
where $W_{n=0}$ and $\varepsilon_{n}$ denote respectively the
superpotential and energy eigenvalue for the unperturbed
potential, $V_{0}=-\ell(\ell+1)\sec{h}^{2}{z}$. Note that the
corresponding wavefunctions reproduce standard properties of the
spherical harmonics \cite{dutt4001997}.

At this stage, with the consideration of Eq. (8)
\begin{equation}
\Delta{W^{2}_{n=0}(z)}-\Delta{W'_{n=0}(z)}+2(\ell\tanh{z)}\Delta{W_{n=0}(z)}=\Delta{V(z)}-\Delta\varepsilon_{n=0},
\end{equation}
one arrives at
\begin{equation}
\Delta{W_{n=0}}=\frac{b}{\ell}~~,
~~~\Delta\varepsilon_{n=0}=-\frac{b^{2}}{\ell^{2}}~~,
~~~\Delta{V(z)}=2b\tanh{z},
\end{equation}
where $b$ is a constant. It makes clear that the full interaction
potential has now the form of the Rosen-Morse II potential. A
brief study of all shape invariant exactly solvable potentials
\cite{cooper2671995} in one dimensional space, together with
\cite{gonul3} related to non-central potentials, clarify the
physically meaningful choice of $b$ as to be $b=-\frac{\gamma}{2}$
with $\gamma$ be another constant relating to the system of
interest. For the completeness, one should now take consider the
whole superpotential,
\begin{equation}
W^{total}_{n=0}(z)=W_{n=0}(z)+
\Delta{W_{n=0}}(z)=\ell\tanh{z}-\frac{\gamma}{2\ell}~~,
\end{equation}
which will reproduce the whole spectrum. Proceeding within the
framework of supersymmetric quantum mechanics, the energy spectrum
for the total potential
$V=-\ell(\ell+1)\sec{h}^{2}{z}-\gamma\tanh{z}$ is given in the
form
\begin{equation}
E_{n}=-(\ell-n)^{2}-\frac{\gamma^{2}}{4(\ell-n)^{2}}=\varepsilon_{n}+\Delta\varepsilon_{n},
\end{equation}
which is the proof of the present theory. To see the exactly
solvable form of the angular potential $U_{2}$, we use inverse
mapping $\sec{h}{z} =\sin\theta,  \tanh{z}=-\cos\theta$ and
bearing in mind Eqs. (14) and (18), then arrives at
\begin{equation}
U_{2}(\theta)=\frac{\gamma\cos\theta}{r^{2}\sin^{2}(\theta)}~~.
\end{equation}
From the mathematical point of view, $r^{2}\sin^{2}\theta$ in the
calculations of exactly solvable forms of $U_{2}$  comes
naturally, see Eqs. (14) and (18). Hence one can generalize the
above potential involving a constant, if necessary, related to the
physical system considered. In this case, such potentials are
given as
\begin{equation}
U_{2}(\theta)=\frac{\beta+\gamma\cos\theta}{r^{2}\sin^{2}(\theta)}~~,
\end{equation}
in which $\beta$ appears, from Eq. (18), only on RHS of Eq. (19)
as a piece of energy value. Considering Eqs. (19) and (25) for the
presence of $\beta$, we obtain
\begin{equation}
\ell=n+\left[\frac{(m^{2}+\beta)+\sqrt{(m^{2}+\beta^{2})-\gamma^{2}}}{2}\right]^{\frac{1}{2}}.
\end{equation}
This $\ell$- value is used in the energy expression given for the
central potential $U_{1}(r)$ to introduce the complete spectrum of
analytically solvable non-central potentials. The present result
agrees with Eq. (10) of Ref.\cite{gonul3}, where the generalized
Coulomb potential was discussed.

Clearly, Eq. (22) is the most significant equation of this Letter,
which defines exactly the possible forms of $U_{2}(\theta)$
yielding analytical solutions. The main point here of course  is
to be able to find an analytical expression for $\Delta{W}$ and
subsequently $\Delta{V}$ for the definition of the total
$\theta$-dependent perturbing potential, which requires exactly
the satisfaction of Eq. (22) leading to a closed form for the
complete spectrum.

One should however note that although we have focussed here on the
eigenvalues, calculation of the corresponding eigenfunctions
within the same model is quite straightforward, see Eq. (2). In
addition, we believe that this generalization would considerably
extend the list of exactly solvable non-central potentials for
which the solution can be obtained algebraically in a simple and
elegant manner as discussed here.

A similar study is conducted here for the shape invariant
P\"{o}sch-Teller II type potential \cite{cooper2671995}, within
the frame of Eqs. (19) through (22). This choice requires
\begin{equation}
\Delta{W_{n=0}}(z)=-\alpha\coth{z},
\end{equation}
which, from (22), reproduces
\begin{equation}
\Delta{V(z)}=U_{2}(z)=\alpha(\alpha-1)\csc{h}^{2}{z}~~~,
~~~\Delta\varepsilon_{n=0}=-\alpha(\alpha-2\ell).
\end{equation}
Thus, the full angle dependent potential in (19) turns into
$V=-\ell(\ell+1)\sec{h}^{2}{z}+\alpha(\alpha-1)\csc{h}^{2}{z}$
having a complete spectrum in the form of
$E_{n}=-(\ell-\alpha-2n)^{2}=\varepsilon_{n}+\Delta\varepsilon_{n}$.
Using the same algebraic procedure as before, one obtains the
related exactly solvable non-central potential as
\begin{equation}
U_{2}(\theta)=\frac{\delta}{r^{2}\sin^{2}(\theta)}+\frac{c}{r^{2}\cos^{2}(\theta)}~~,
\end{equation}
in which $\delta$ and $c=\alpha(\alpha-1)$ are constants. As
explained in the previous example, $ \delta $ appears as a piece
of energy eigenvalue on RHS of Eq. (19), therefore
\begin{equation}
m^{2}+\delta=(\ell-\alpha-2n)^{2}
\end{equation}
from which one defines the $\ell$ - value as
\begin{equation}
\ell=2n+\left(\frac{1}{2}\pm\sqrt{\frac{1}{4}+c}\right)+(m^{2}+\delta)^{\frac{1}{2}},
\end{equation}
that is the same result with compared to (26) of \cite{gonul3},
and also agrees with the related references in \cite{gonul3}. If
$U_{1}(r)$ is taken as the harmonic oscillator potential, then the
whole non-central potential with $U_{3}(\varphi)=0$ corresponds to
the generalized oscillatory potential. To find the full spectrum
for such a potential, Eq. (33) is invoked to the energy spectrum
of $U_{1}(r)$ \cite{gonul3}.
\section{Concluding Remarks}
In this Letter, we have attempted to explore the effectiveness of
the recently developed formalism through which we have made
successfully the complete mathematical analysis of the reason
behind exact solvability of some Schr\"{o}dinger equations with a
class of non-central but separable potentials, for which the
complete spectrum and eigenfunctions can be written down
algebraically using the well known results for the shape invariant
potentials. Generalization of our technique to other non-central
potentials is quite straightforward and  use of the present model
may also be useful for solving other quantum mechanical
complicated systems analytically. With the above consideration the
authors hope to stimulate further examples of applications of the
model in important problems of physics, which requires further
technical work that will be discussed in a forthcoming paper.

\end{document}